\documentclass[review,12pt, 3p]{elsarticle}

\usepackage{amssymb}

\usepackage{graphicx}
\graphicspath{
	{images/}
}

\usepackage{subfiles}

\usepackage{upgreek}

\usepackage[hyphens]{url}

\usepackage{amsmath}
\usepackage{siunitx}

\usepackage{textcomp}
\usepackage{listings}
\journal{Digital Investigation}

\usepackage{subfig}

\usepackage{xcolor}

\begin{document}
	
	\begin{frontmatter}



\title{Reverse Engineering the Raspberry Pi Camera V2: A study of Pixel Non-Uniformity using a Scanning Electron Microscope\tnoteref{copyright}}


\author[eee]{R.~Matthews\corref{cor1}}
\ead{richard.matthews@adelaide.edu.au}
\author[eee]{M.~Sorell}
\author[cs]{N.~Falkner}
\cortext[cor1]{Corresponding author}
\tnotetext[copyright]{ Copyright 2018. This manuscript version is made available under the CC-BY-NC-ND 4.0 license http://creativecommons.org/licenses/by-nc-nd/4.0/}

\address[eee]{The University of Adelaide, School of Electrical and Electronic Engineering, Adelaide, SA, 5005 AUS.}
\address[cs]{The University of Adelaide, School of Computer Science, Adelaide, SA, 5005 AUS.}

\begin{abstract}
In this paper we reverse engineer the Sony IMX219PQ image sensor, otherwise known as the Raspberry Pi Camera v2.0. We provide a visual reference for pixel non-uniformity by analysing variations in transistor length, microlens optic system and in the photodiode. We use these measurements to demonstrate irregularities at the microscopic level and link this to the signal variation measured as pixel non-uniformity used for unique identification of discrete image sensors. 

\end{abstract}


\end{frontmatter}



\section{Introduction}
Sensor pattern noise (SPN) has been accepted as a viable method to identify a unique camera responsible for taking a specific image. These methods rely on the non-uniform nature of individual pixels, known as pixel non-uniformity (PNU), to establish a unique reference pattern. There exists a gap in the literature which clearly explains the cause of PNU especially with the purpose of instructing jurors in mind. Sensor pattern noise (SPN) methods based on PNU are important for intelligence communities, law enforcement communities and can have additional applications for photographers wishing to establish ownership without relying on metadata or additional watermarks. While SPN methods are accepted by the forensic community (an important Daubert criterion) there is a risk that a lay juror with no mathematical background may be confused by the evidence presented in a court setting. This can be confounded by issues such as the CSI effect where a juror expects Hollywood science to replace real forensic methodology.

In this paper we explore some of the physical characteristics of the IMX219PQ image sensor, providing direct evidence of variation which may cause SPN. We reverse engineer the design, and analyse three features of an image sensor under a scanning electron microscope: transfer gate length, variations within the microlens optic system (LOS) and variations within the photodiode region. It is hypothesised that the variations within these features are principally caused by tolerances in the manufacturing process, can visually be seen under significant magnification and may cause PNU.

\section{Related Work}\label{RelatedWork}

	The existing literature provides a blanket statement for the source of PNU \cite{lukas2006digital}. ``PNU is caused by the inhomogeneity of silicon wafers and other imperfections during the manufacturing process.'' The focus of the literature since has been directed on improving the method first identified as distinct from identifying the underlying source. A gap exists to address this underlying assumption as to the source of PNU within the pixel unit of an image sensor. Correcting this assumption provides greater clarity for forensic identification purposes should the literature be challenged in a legal setting. 
	
	We focus on three feature categories which may cause the PNU used to uniquely identify each sensor. These are variations within the length of the transfer gate, within the micro lens optical system (LOS) and within the photo diode itself. The justification for this approach is given in the related works below.
	
	\cite{Rizzolo2018} provides an analysis of key geometric properties within the structures of an image sensor and how this affects the charge transfer efficiency (CTE). Variable CTE values for discrete pixels will cause PNU. The geometric aspects investigated by \cite{Rizzolo2018} include the photodiode size, the transfer gate length and the sense node (SN) storage area. The SN is also referred to as a floating diffusion (FD). The two terms are able to be used interchangeably. In both cases, the region refers to a highly-conductive region without resistive connection to allow the storage of charge, in effect a capacitor. \cite{Rizzolo2018} showed that in all instances of geometric variation it was the length of these critical geometries which affected the CTE of the device. Incomplete transfer between photodiode and SN also causes image lag and noise on an individual pixel \cite{Fossum2104}. This provides justification to measure the lengths of the transfer gate as well as the photo diode as a possible source of PNU. 
	
	Differences in the transfer gate are also responsible for dark current variations. Dark current is generated in the transfer gate by the silicon to silicon dioxide interfaces or by defects below the surface of the gate itself \cite{Fossum2104}. We have previously shown the use of dark current for thermal identification of image sensors \cite{Matthews2018Thermal, matthews2018temperature}. Dark current has also been used in isolation and in hybrid methods for image identification alongside PRNU \cite{kurosawa2013casestudies}. Measuring differences within the transfer gate thus provides additional motivation as larger gates will provide greater surface area interactions for dark current generation. 
	
	We have previously shown the contamination of lens effects within the SPN methodology \cite{Matthews2017Isolation, Matthews2018Isolation, matthews2018analysis}. In \cite{chandrakantanon2017} the non-uniform output of a photodiode was presented. \cite[Fig. 2 c]{chandrakantanon2017} shows the non-uniform output of the diode while being excited by a 7V laser through various incident angles. The output of the photodiode was measured to vary between 305mv and 208mv depending on the incident angle of the laser. Given the non-uniform output demonstrated in this work it is hypothesised that this could be a contributor to PNU in image sensors. Assuming this is a contributor of PNU in image sensors, it is theorised that this would manifest via misalignment within the micro LOS of a sensor focusing or filtering photons to different areas of the depletion region of the photodiode. This would be distinctively different to PNU being caused by variance in doping levels within the photodiode or the size of the photodiode during manufacture. Such misalignments should be possible to visually see under cross-sectional analysis through the use of a scanning electron microscope. This provides justification to examine the micro LOS of the image sensor.

\section{Research Methodology}\label{materials}

\begin{figure}[!ht] 
		\centering
		\subfloat[][]{	
			\includegraphics[width=7.5cm]{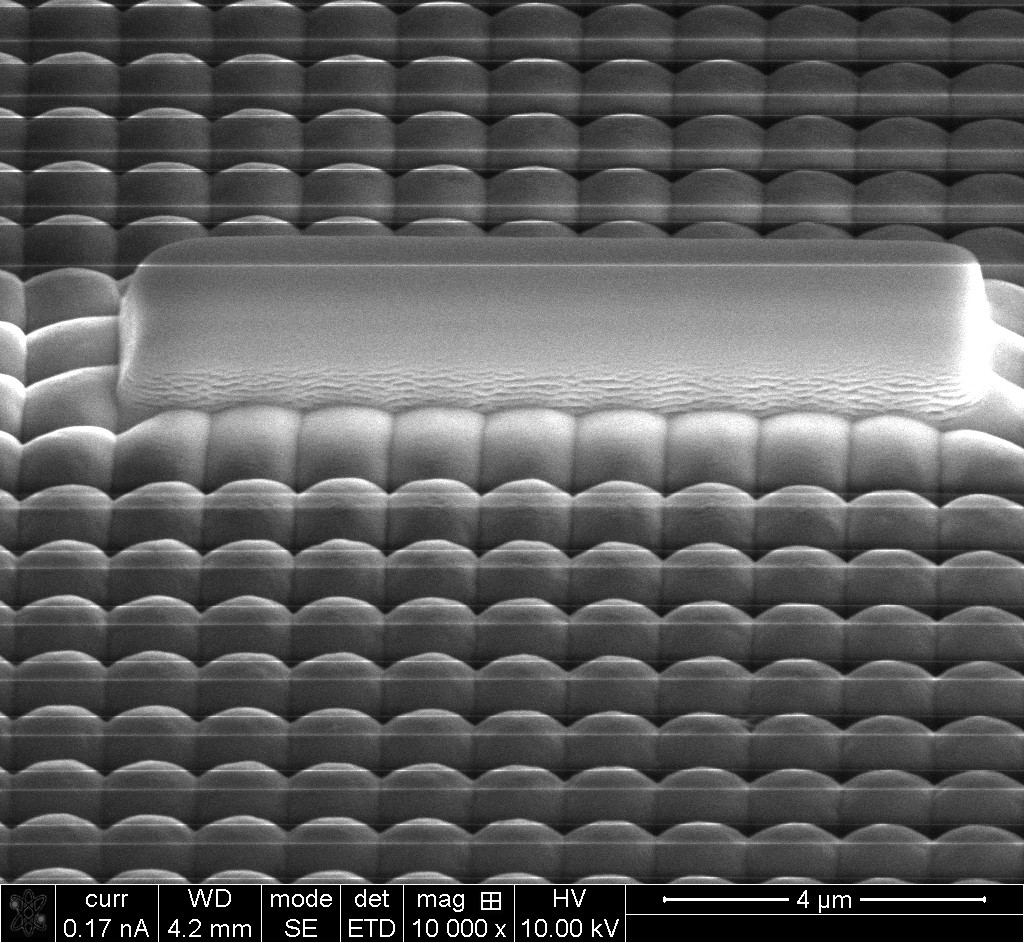}
		}
		\subfloat[][]{	
			\includegraphics[width=7.5cm]{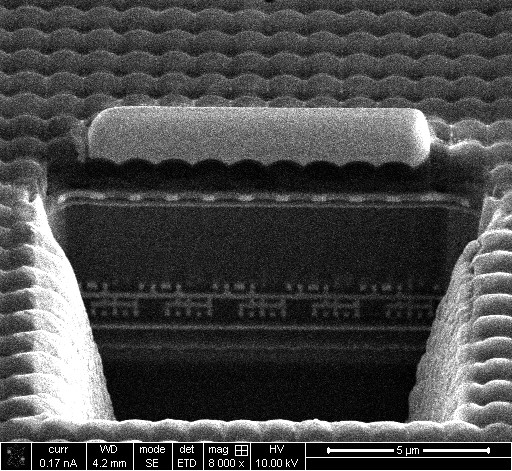}
		}
		\caption[SEM Microscope Excavation Process]{(a) Platinum (shown here as a growth on top of the micro-array) is deposited on the CIS to prevent micro-fractures forming during the excavation process. (b) The excavated region of the CIS is shown after the Gallium ion beam has been used to step out the material present in the region of interest. Several passes are used to obtain a smooth, polished cross-section.}
		
		\label{Platinum}
	\end{figure}

	Using a Sony IMX219PQ \cite{sonyIMX219} image sensor as the test subject imaging was performed under a FEI Dual Beam Helios Nano Lab 600 scanning electron microscope \cite{fei}. The Helios allows imaging down to 1nm at 15kV. Minimal preparation of the sample is required before imaging since the image sensor is made primarily of silicon and hence, is conductive to the electron beam. The IMX219PQ comes as a package board from the Raspberry Pi Foundation. The sensor is de-constructed to obtain the silicon wafer in isolation to the peripheral circuit and attached to the FEI imaging platform by way of adhesive conductive dot. The platform is tilted to 52° to enable imaging and machining by the gallium focused ion beam. Under magnification, a layer of platinum was deposited above the area to be excavated to prevent fracturing (Figure \ref{Platinum}{a}).  After the platinum is deposited a process of staged cuts were made using a gallium ion beam to create an excavated area through the sensor that can be imaged (Figure \ref{Platinum}{b}). The gallium ion beam is capable of machining down to 5nm allowing precise cuts to be made. This is a destructive process.

	Using a magnification of up to 100,000x (400nm scale), a current of 0.086 - 0.17nA and voltage of 5-15kV, images are then obtained of the top layers of the integrated CIS containing the pinned photodiode and the associated supply and readout circuitry. Only the top silicon wafer is studied.

\section{Data Collection and Analysis}\label{results}

	\subsection{Reverse Engineering}
\begin{figure}[h!] 
			\centering
			\includegraphics[width=0.8\textwidth]{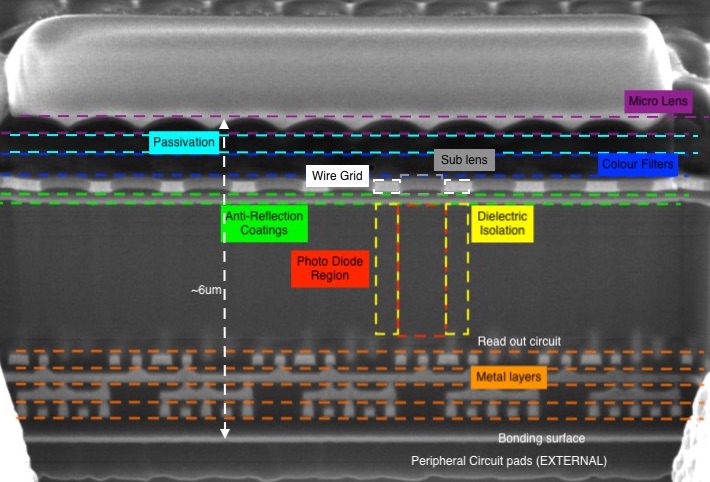}
			\caption{A horizontal cross-section of the IMX219PQ sensor. Each region of the sensor is clearly identified.}
			\label{horizontalcross}
		\end{figure}

\begin{figure}[h!] 
			\centering
			\includegraphics[width=0.8\textwidth]{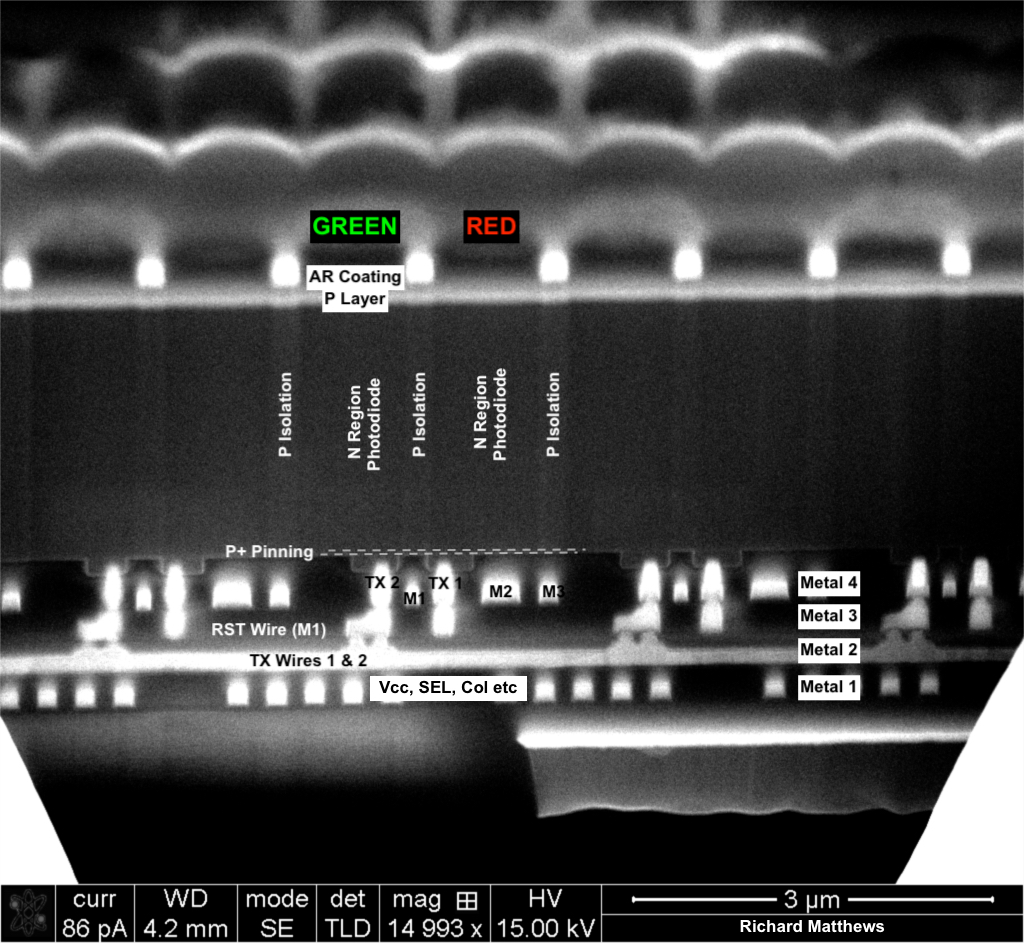}
			\caption{A secondary cross-section of the IMX219PQ sensor. The structure of the photodiode and circuit is identified within the image.}
			\label{Dechorizontalcross}
		\end{figure}

		Using \cite{Fossum2104}, Figure \ref{horizontalcross} and \ref{Dechorizontalcross} presents the layers of the CIS with each region identified. The sensor is a back illuminated sensors (BIS) with a pinned photodiode (PPD). The P+ pinning layer can be seen in Figure \ref{Dechorizontalcross}. In Figure \ref{horizontalcross} the width of the sensor is measured as approximately 6um. This is consistent with the provided literature from Sony \cite{sonyIMX219Design} which also indicates a dual wafer design. This is confirmed in these images with the bonding surface between the two wafers indicated in Figure \ref{horizontalcross} at the appropriate distance from the top of the micro LOS. In the micro LOS the structure can be seen as a main lens, passivation layer, Bayer filter and then a minor sub lens nestled in between the nodes of the wire grid. The photodiode region is identified and isolated between two layers of P doped semiconductor used to prevent cross talk. The isolation regions are located directly beneath the wire grid used for optical isolation. This is seen clearer in Figure \ref{Dechorizontalcross} where the microscope is focused onto the read out circuit layer of the image sensor.
			 
		Using reference images from \cite{Fossum2104}, the Bayer filter elements visible in Figure \ref{Dechorizontalcross} are identified as green and red. This is made out as the red filter elements are thinner than green with the blue elements being thicker and filling almost the entire layer. The transfer nodes (TX) are  visible within the image. From here it is possible to identify the individual transfer, M1 reset (RST), M2 source follower (SF) and M3 column select transistors within the Metal 4 layer. The wiring for each of these nodes is then routed to the underlying metal layers to provide read out access to the circuit, most likely on the peripheral or underlying silicon wafer. The TX wires are seen within the Metal 2 layer while it is assumed that the Vcc, select and column bus tracks are visible in Metal 1. 
				
\begin{figure}[h!] 
			\centering
			\includegraphics[width=0.8\textwidth]{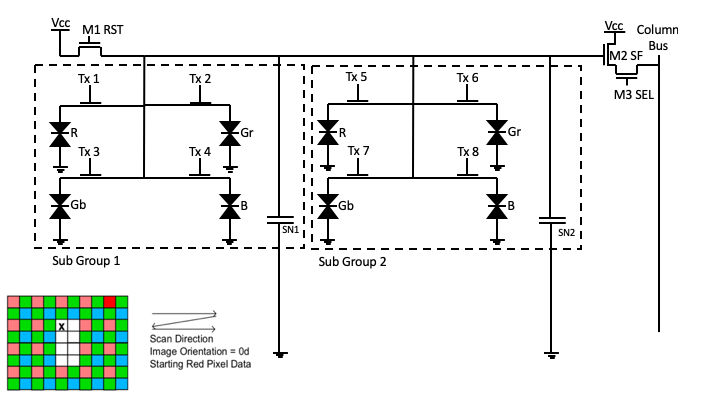}
			\caption{The equivalent circuit for the IMX219PQ pixel unit. Inset shows the scanning direction for read out of the IMX219PQ sensor when identifying a 2x4 pixel defect on the sensor taken from \cite{sonyIMX219Full}. }
			\label{EquivilentCircuit}
		\end{figure}

			Using the information identified above in conjunction with  \cite{sonyIMX219Full} and \cite{Fossum2104} an equivalent circuit can be identified for the Sony IMX219PQ BIS pixel unit. From examining \cite{sonyIMX219Full} we see that the sensor has the capability to store defective pixels in one of three modes: single pixels, single adjacent pixels by Bayer element and individual blocks of 2x4 adjacent pixels. Understanding how the sensor stores defects provides insight to the readout structure. It follows that the sensor has a pixel unit comprising of 2x4 photodiodes sharing a single readout circuit. Using \cite{Fossum2104} the pixel unit is comprised of a circuit of eleven transistors for every eight photodiodes. Using the familiar nomenclature of average number of transistors per pixel we see the IMX219PQ is a 1.375T (11 transistors per 8 pixels) design. This is indicated in Figure \ref{EquivilentCircuit}. 
		
		We suspect that the clover leaf pattern is utilised as per previous Sony iterations, in particular the sensor (Sony IMX145) in \cite{Fossum2104}. Unlike the equivalent circuit shown in \cite[Fig. 9]{Fossum2104} we show only two sense node locations corresponding to the sense node in each of the two pixel subgroups noting the work of Fontaine in \cite[Fig. 10]{Fossum2104} showing these two locations in the centre of the clover leaf. We note that the charge for each photodiode when read out will be distributed on both of these nodes in parallel due to Kirchoff's Current Law. The readout circuitry is an important distinction to clarify as it forms a significant source of dark current. Sensors that demonstrate a different readout structure (3T, 4T, 2T, 1.75T) provide a possible source of difference for forensic identification. The dark current of such sensor configurations should be measured in future work. The clover leaf pattern should also be confirmed by imaging the bottom of the sensor.

	\subsection{Transfer Gate} \label{TransferGate}
		Measurements were taken of the readout node of the photo diode using the SEM built in measurement tool. These were then verified manually by printing images to scale and measuring by hand. These results are shown in Table \ref{TransferGateSize}.
		
		\begin{figure}[!ht] 
			\centering
			\subfloat[][]{	
				\includegraphics[width=5cm]{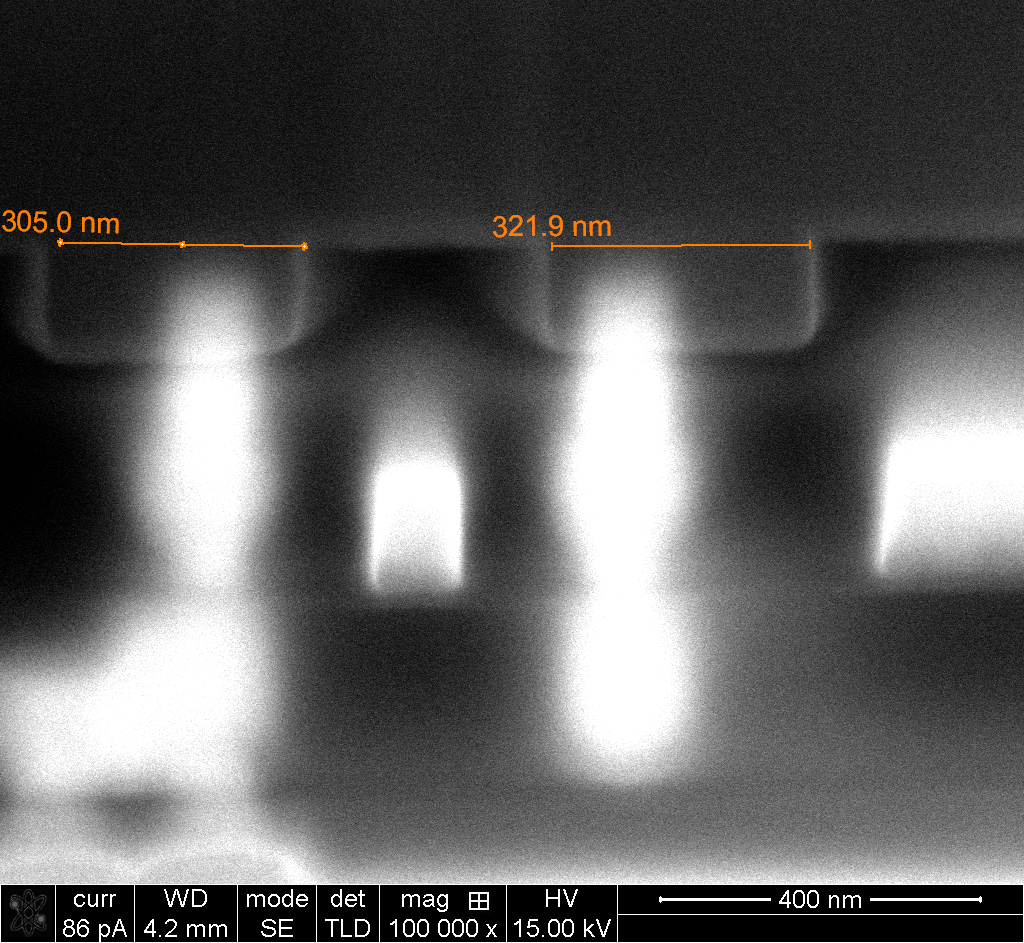}
			}
			\subfloat[][]{	
				\includegraphics[width=5cm]{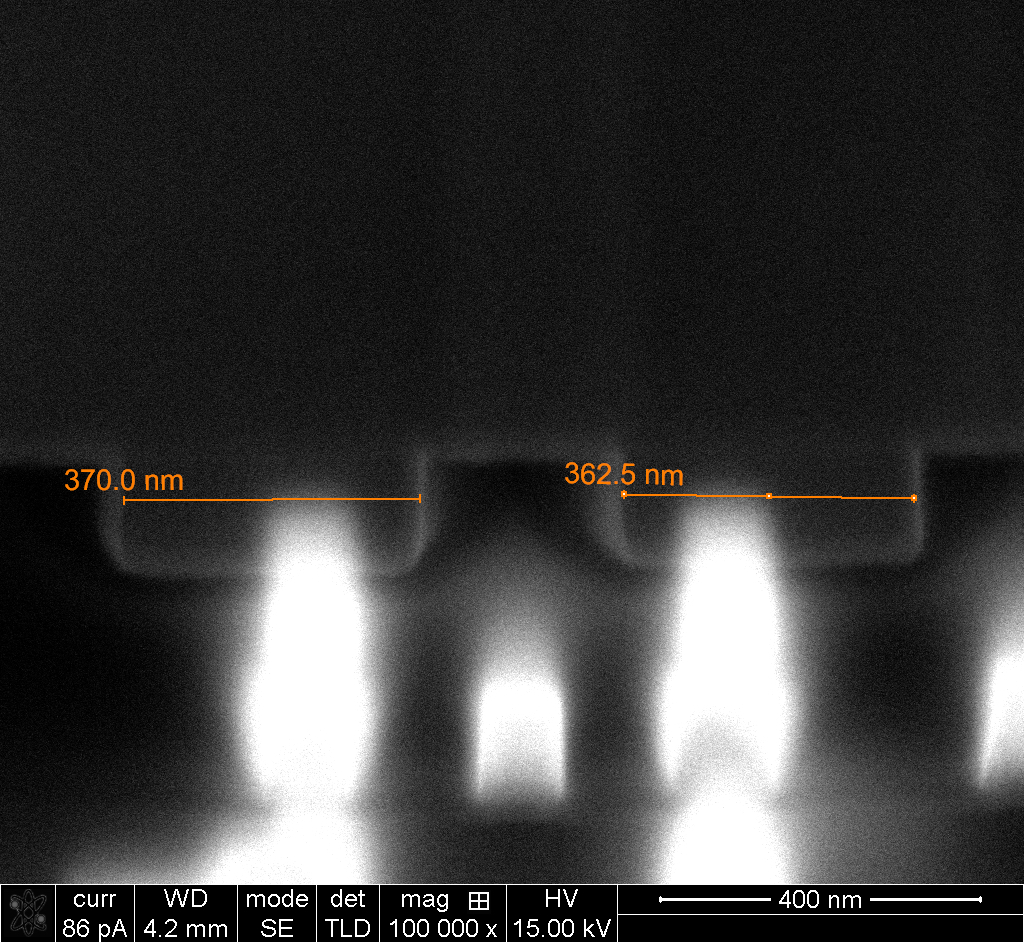}
			}
			\subfloat[][]{	
				\includegraphics[width=5cm]{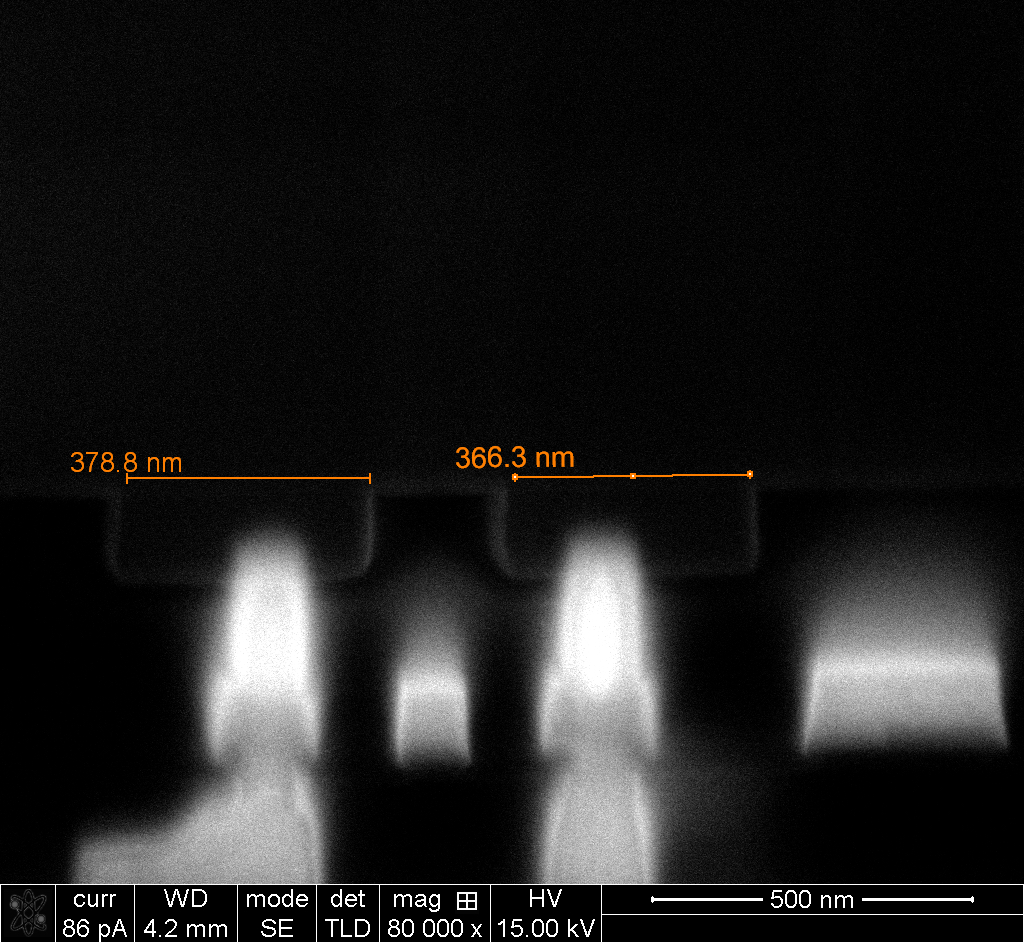}
			}
			\caption[]{The length of six transfer gates are taken across three separate pixel units Odd gates are green, even gates are red. (a) Transfer gates 1 and 2. (b) Transfer gates 3 and 4. (c) Transfer gates 5 and 6.}
			
			\label{TransferGates}
		\end{figure}
		
		\begin{table}[h!] 
			\centering
			\caption{Transfer Gate Length} 
			\begin{tabular}{cccc} 
				\hline
				Pixel & SEM Measure (nm) & Manual Measure (nm) & Average (nm)\\ 
				\hline
				1    & 305 & 308 & 307 \\
				2    & 322 & 329 & 326 \\
				3    & 370 & 370 & 370 \\
				4    & 363 & 364 & 364 \\
				5    & 379 & 380 & 380 \\
				6    & 366 & 378 & 372 \\
				\hline
			\end{tabular}
			\label{TransferGateSize}
		\end{table}
	
		There is variation with all transfer nodes measured. The mean of all six transfer nodes is 353nm. Measuring the variance to the mean from each transfer node we obtain a double sided tolerance of +27nm and -46nm. Extrapolating, this leads to a feature size of 350nm with an engineering tolerance of +/-50nm. The polysilicon layer has an average thickness of 160nm with metal 4 being approximately 280nm in thickness. This is consistent with the 130nm lithography process design rules in  \cite{Tyagi2000} noting we have labelled our metal layers in reverse order. This measurement is not consistent throughout the layer with the layer increasing and decreasing in thickness throughout the images obtained. This thickness variation will cause minor geometric changes to the features of the circuit. These thickness variations will become more pronounced in top layers as the variations stack.

	\subsection{Micro Lens Optical System} \label{microLOS}
		
\begin{figure}[!ht] 
			\centering
			\subfloat[][]{	
				\includegraphics[width=7.5cm]{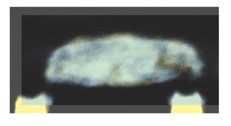}
			}
			\subfloat[][]{	
				\includegraphics[width=7.5cm]{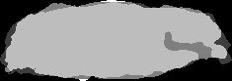}
			}
			\caption[SEM Microscope Excavation Process]{(a) Overlay of the two Green-Red Bayer filter elements. (b) The overlay of the two Green-Red Bayer filter elements as a binary image using FIJI \cite{FIJI}.}
			
			\label{BayerOverlay}
		\end{figure}

		Two Green-Red Bayer elements are overlaid in Figure \ref{BayerOverlay}. In \ref{BayerOverlay}{a} the microimagery is presented for the two immediate, same colour, neighbouring elements with noticeable variations in the structure of these two identical Bayer elements. The stained blue shows a large segment missing to the right where as the stained yellow is slightly thicker. The cropped images are run through a Canny edge detector in FIJI \cite{FIJI} to provide a binary gate to compare the two filter elements against. This result of the Canny edge detector is presented in \ref{BayerOverlay}{b}. White is where element 1, previously tinted yellow, is presented. Dark grey is element 2 previously tinted blue. Dark white is where both elements are, and light grey is where only element 1 is present. This shows the contrasting differences between the two CFA elements with a  hole in element 2 positioned above the optical block (metal grid). These minor variations are likely to create optical aberrations known as Seidel aberrations affecting where individual photons will strike the depletion region of the photodiode \cite{Seidel_1857,jenkins1957fundamentals}.
		
\begin{figure}[h!] 
			\centering
			\includegraphics[width=0.8\textwidth]{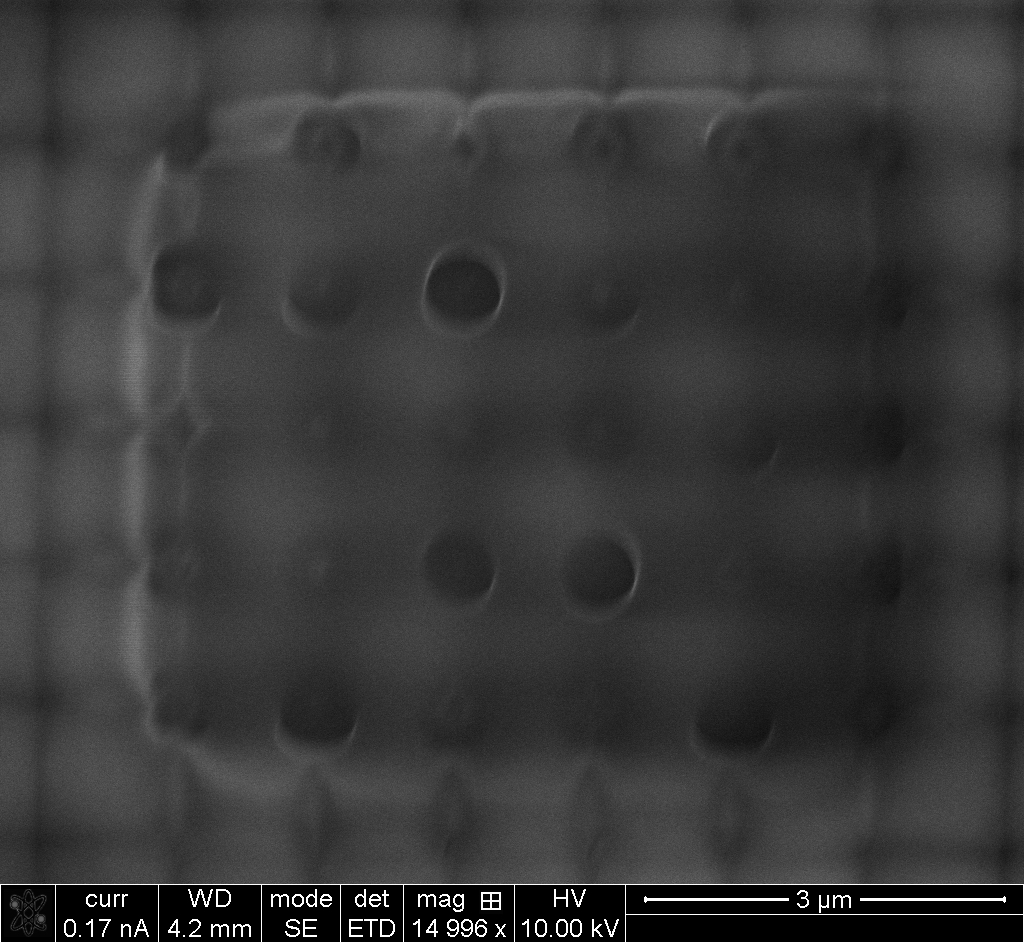}
			\caption{Sectioning the top lens layer of the IC reveals the passivate later above the Bayer elements below the micro lenses. Bubbles are evident above the wire grid in random locations on the surface of the chip.}
			\label{WireGridBubbles}
		\end{figure}

		The Bayer filter is not the only region that is observed to have irregularities which will cause optical aberrations. Sectioning off the top layer of the micro LOS provides access to the passivation layer in between the micro lens and the Bayer elements.  Shown in Figure \ref{WireGridBubbles} holes in the passivation layer are visible directly above the wire grid elements. This is another defect within the structure of the sensor that is not uniformly distributed on every pixel. Once again, the formation of holes within the upper layers of the micro LOS are likely to cause aberrations \cite{Seidel_1857,jenkins1957fundamentals}.
	
	\subsection{Photodiode Variations} \label{PDVariation}
	
\begin{figure}[h!] 
			\centering
			\includegraphics[width=0.8\textwidth]{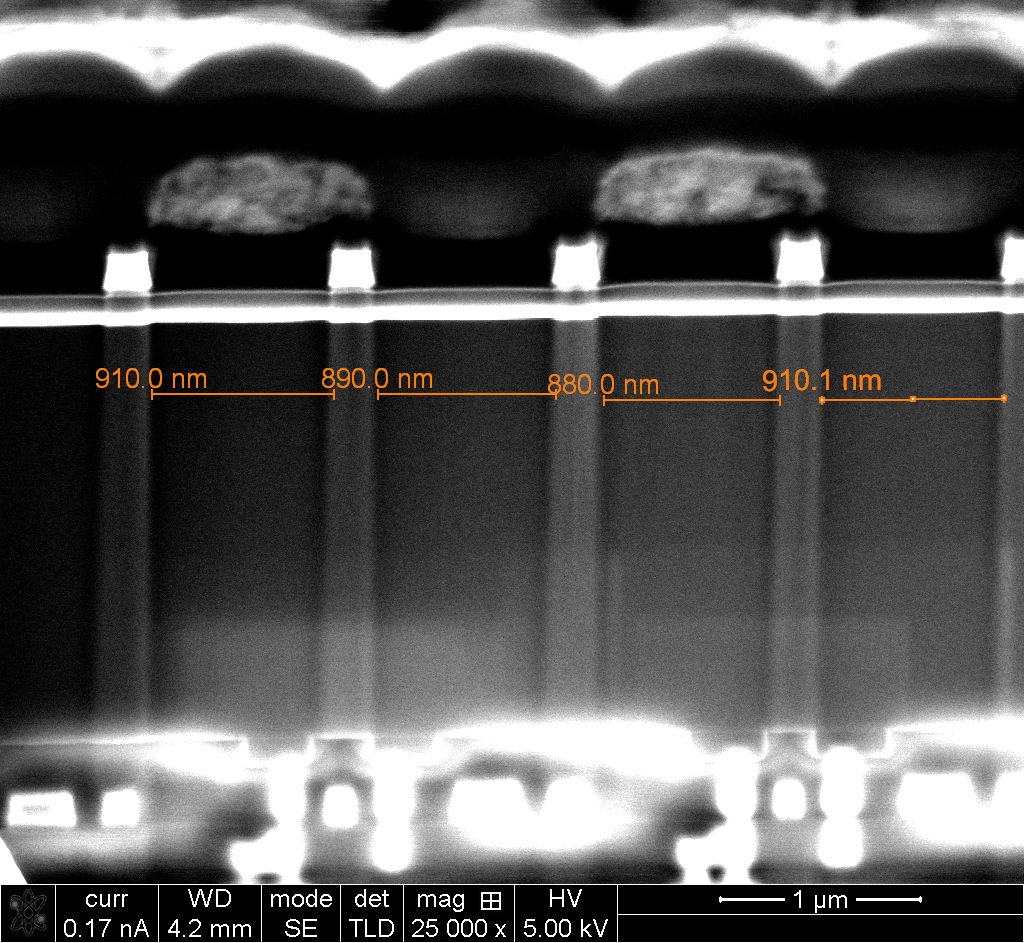}
			\caption{Four photodiodes from two separate pixel units compared for length.}
			\label{PhotoDiodeVariation}
		\end{figure}

		Four photodiodes and their length are shown in Figure \ref{PhotoDiodeVariation}. We note the presence of a charging artefact due to the highly conductive metal layers causing a smearing effect into the photodiode region on the image. Measurements are taken well away from this artefact to avoid any contamination. Measurements are taken from the SEM and then confirmed via manual scaling as per section \ref{TransferGate}. These measurements are displayed in Table \ref{PhotodiodeLength}. All measurements show variation from photodiode to photodiode.
		
		\begin{table}[h!] 
			\centering
			\caption{Photodiode length} 
			\begin{tabular}{cccc} 
				\hline
				Photodiode & SEM Measure (nm) & Manual Measure (nm) & Average (nm)\\
				\hline
				1 Gr   & 910 & 899 & 905 \\
				2 R    & 890 & 873 & 882 \\
				3 Gr  & 880 & 879 & 880 \\
				4 R   & 910 & 905 & 908  \\
				\hline
			\end{tabular}
			\label{PhotodiodeLength}
		\end{table}
		
		The mean photodiode length is measured as 894nm with a variance of +/-14nm. The nominal pixel length is stated as being 1120nm (1.12um) \cite{sonyIMX219Full}. Measuring the length of the isolation in between the photodiode regions we can obtain a measurement for the average pixel length.  The P dielectric isolation is manually measured as above and displayed in Table \ref{IsolationLength}. A mean measurement is obtained as 240nm. Adding the two mean measurments together obtains an average pixel length of 1134nm (1.13um). The four pixels in Figure \ref{PhotoDiodeVariation} are manually measured to obtained a length of 451nm. This provides a mean measurement of 1127.5nm (1.13um). These measurements indicate that the pixel length is slightly larger than stated in the documentation due to manufacturing tolerance.

		\begin{table}[h!] 
			\centering
			\caption{Isolation Length} 
			\begin{tabular}{cc} 
				\hline
				P Isolation &Manual Measure (nm) \\
				\hline
				1 Gr   & 232  \\
				2 R    & 234  \\
				3 Gr  & 264  \\
				4 R   & 228   \\
				\hline
			\end{tabular}
			\label{IsolationLength}
		\end{table}
		
		The variance in sizes across pixels are likely to be a significant cause of PNU on the sensor as different sized photodiodes will measure different amounts of light and have different amounts of dark current.

\section{Discussion}\label{discussion}

Dark current and Photo Response Non-Uniformity noise are leading candidates in SPN methods for identification of discrete image sensors from a candidate photo. \cite{kurosawa2013casestudies} and \cite{lukas2006digital} are both indications of how this method can be used with great success. The defects shown in Sections \ref{TransferGate} and \ref{PDVariation} are the features within the image sensor and the electrical circuit which are likely candidates to cause the non-uniform characteristics of both Dark Current and Photo Response non-uniformity used for unique identification. 

Lens aberrations are known to link images to cameras \cite{san2006source}. Shown in Section \ref{microLOS} are irregularities in the micro LOS that is inseparable from the image sensor. Lens aberrations normally link an image to a LOS that is able to be separated from a camera. In the case presented here, the defects within the micro LOS are inseparable from the sensor itself and as such should provide a forensic link to the image sensor as opposed to a LOS. 

It is important to note that when we and the literature have referred to manufacturing defects we are not referring quality control issues in the typical sense. Where defects would normally cause the scrap of the sensor or entire wafer, the defects we discuss are minimal and still allow the sensor to operate within the designed engineering tolerance. It is this variability within tolerance that is being exploited here for forensic identification. While this trace is treated as unique it is also important to note that this has not been proven. The largest scale test of this method has been done by \cite{goljan2009large} where it was found, using the current state-of-the-art method, PCE identification has a false acceptance rate of less than 3 in 125,000. As this work has been conducted with the juror in mind, it is important to highlight this distinction for the purposes of the Daubert criterion.

\section{Conclusion and Future Work} \label{conclusion}

In this paper we have demonstrated pixel non-uniformity within the silicon structure of the image sensor at the microscopic level. We have demonstrated discontinuities between discrete layers within the image sensor. While these discontinuities do not affect the overall image performance capabilities of the sensor they do contribute to a layer of additive noise known as PNU. Particular attention has been paid to the transfer gates and the associated read out circuitry. This work has drawn attention to the variability of physical characteristics of the electronic circuits on an imaging sensor to visualise how sensor pattern noise may, at least in part, be explained. We have not considered the variation in semiconductor performance due to chemical-level variability such as doping concentration and contamination, nor have other read-out circuitry configurations been analysed, which are matters for future study. Our exploration of the physical dimensional variability may also be useful in explaining, in part, the origins of sensor pattern noise to a lay audience

\section{Acknowledgements}
This research did not receive any specific grant from funding agencies in the public, commercial, or not-for-profit sectors. This work was supported with supercomputing resources provided by the Phoenix HPC service at the University of Adelaide. The authors acknowledge the facilities, the scientific and technical assistance, of the Australian Microscopy and Microanalysis Research Facility at Adelaide Microscopy, the University of Adelaide. This research is supported by an Australian Government Research Training Program (RTP) Scholarship and forms part of a thesis chapter.




\bibliographystyle{elsarticle-num} 
\bibliography{ELS-Microscopy}

\end{document}